# Associations between learning assistants, passing introductory physics, and equity: A quantitative critical race theory investigation

Ben Van Dusen[1] and Jayson Nissen[2]

[1]*Department of Science Education, California State University Chico, Chico, California 95929, USA*
[2]*J. M. Nissen Consulting, Corvallis, Oregon 97333, USA*

Many science, technology, engineering, and math degrees require passing an introductory physics course. Physics courses often have high failure rates that disproportionately harm students who are historically and continually marginalized by racism, sexism, and classism. We examined the associations between learning assistant (LA) supported courses and equity in nonpassing grades [i.e., drop, fail, or withdrawal (DFW)] in introductory physics courses. The data used in the study came from 2312 students in 41 sections of introductory physics courses at a regional Hispanic serving institution. We developed hierarchical generalized linear models of student DFW rates that accounted for gender, race, first-generation status, and LA-supported instruction. We used a quantitative critical race theory (QuantCrit) perspective focused on the role of hegemonic power structures in perpetuating inequitable student outcomes. Our QuantCrit perspective informed our research questions, methods, and interpretations of findings. The models associated LAs with overall decreases in DFW rates and larger decreases in DFW rates for Black, Indigenous, and people of color than their White peers. While the inequities in DFW rates were lower in LA-supported courses, they were still present.

DOI: 10.1103/PhysRevPhysEducRes.16.010117

## I. INTRODUCTION

Introductory physics courses are required for many science, technology, engineering, and math (STEM) degrees. The high rates of students earning nonpassing grades [i.e., drop, fail, or withdrawal (DFW)] in these courses creates a significant barrier to many students earning a STEM degree [1–5]. For many students, other responsibilities (e.g., family or financial) make postponing their graduation to retake a course an impossibility, leading them to either change majors or drop out of school entirely [6,7]. These negative impacts are particularly detrimental to students whose success is already suppressed by institutional racism, sexism, and classism [7,8]. Scholars have found overwhelming evidence for the role of racism and sexism in physics education. While women perform as well or better than men in their STEM courses, they disproportionately leave because of the hostile and competitive environment they experience in STEM courses [9]. Most women report experiencing sexual harassment in their physics education [10,11]. Black, Indigenous, and people of color (BIPOC) face a similarly harsh learning environment in physics. Faculty and mentors advise BIPOC students of color against pursuing STEM degrees; their peers and instructors avoid or ignore them and deny them insider knowledge for success (e.g., courses to take or who to ask for assistance) [9,12–18]. Intersectionality studies tell us that these burdens fall disproportionately on students with multiple intersecting historically marginalized [19] identities [20–22]. For example, Dortch and Patel [12] found that Black women were marginalized by White [23] men, White women, and Black men at both primarily White institutions and historically Black colleges and universities. While racism is experienced differently across marginalized racial groups [24], similar patterns of intersecting oppressive educational power structures have been identified as marginalizing all BIPOC students [25–29]. By disproportionately filtering out students from marginalized groups, physics courses decrease access to and diversity in STEM degrees and the personal and professional opportunities they afford.

Physics faculty integrate learning assistants (LAs) into their courses to support implementing evidence-based teaching strategies that increase learning and can lower DFW rates. The LA model is not a research-based pedagogy, rather it provides structure to support instructors in implementing research-based pedagogies that fit their curricular needs [30,31]. In the LA model, institutions hire undergraduate students to support instructors using evidence-based, student-centered learning practices in their courses. The LA model trains the LAs in pedagogical knowledge and pedagogical content knowledge [32,33] to support them as effective near-peer educators in the classroom. While LAs have been associated with lower DFW rates [34,35], little research has examined the relationships between LA programs and DFW rates for students from marginalized groups.

In this study, we set out to investigate the associations between LAs and equity in DFW rates using data from California State University (CSU) Chico, a rural teaching-intensive Hispanic serving institution. CSU Chico represents an underrepresented institutional context in the literature, which overrepresents research intensive institutions that disproportionately serve white, upper-middle class students [36]. Besides broadening the institution-type representation, the diversity in student population at CSU Chico supports our ability to quantitatively examine outcomes for students from marginalized groups. We use this work to make recommendations through an antiracist lens to combat racism, sexism, and classism in physics education. To accomplish this goal, we analyze and interpret our data through a conceptual framework grounded in quantitative critical race theory (QuantCrit) [37].

## II. LITERATURE REVIEW

There are a limited number of quantitative studies published on equity in physics education. What does exist has largely not examined issues of race and we know of no studies that examined outcomes for gender nonconforming, gender nonbinary, trans, or two-spirit physics students. In our literature review, we tried to highlight contributions from authors from historically marginalized groups.

The dynamic nature of the LA model has enabled instructors using LAs to achieve a wide range of goals. Researchers have investigated the association between LAs and student learning [38,39], course transformation [40], departmental transformation [41,42], teacher recruitment [30], teacher preparation [43,44], equity [45,46], supporting students beyond learning [47,48], and supporting novel learning environments [49,50]. While the association between LAs and student DFW rates interests many constituents [51], we only know of three publications that have examined it [34,35,52].

Alzen *et al.* [52] examined the association between LAs and DFW rates in introductory physics courses for 4941 students at the University of Colorado Boulder over a 10-year period. They analyzed the data using logistic regression models. The variables included in their final model were gender, race, first-generation status, instructor, financial aid, high school GPA, credits at entry, admission test scores, LA exposure, and a placeholder variable for the instructor with no interaction effects between the variables [53]. Their model found that students in LA-supported courses had lower DFW rates than their peers in non-LA-supported courses. The positive effects were present for all student demographic groups, with the white men having the smallest predicted decrease (4%–6%) and first-generation women having the largest predicted decrease (10%–14%). They note, however, that their models did not account for the students in the LA-supported courses overrepresenting women and having better preparation, more college credits, higher high school GPAs, and higher admissions test scores than students in non-LA-supported courses.

In a followup investigation, Alzen *et al.* [34] expanded their dataset to include 32 071 University of Colorado Boulder students across multiple STEM disciplines over 16 years. This investigation accounted for additional student attributes, such as high school grade point average, credits at entry, and admission test scores. In creating their logistic regression models, they only included statistically significant variables. This led to the removal of interaction terms between LAs and all demographic variables other than gender. Their final model predicted that, after accounting for prior preparation, LA support predicted lower DFW rates across all demographic groups. Results showed that men had higher DFW rates than women and that DFW rates were higher for nonwhite students and first-generation students. Unlike their prior investigation, they found that men had larger predicted decreases in DFW rates in LA-supported courses than women. Both of Alzen *et al.*'s studies [34,52] were situated in the same institutional context, limiting the generalizability of their findings. Specifically, the University of Colorado Boulder is a research intensive university and while it is a public institution, 41.7% [54] of the students come from out of state and only 15.7% of student receive Pell Grants compared to 32% nationally. The University of Colorado Boulder is also the institution that developed the LA model. The students and instructors in the study had different resources to support their success than many of their peers in other institutions with LA programs.

Close *et al.* [35] examined the associations between LAs and introductory physics student outcomes at Texas State University San Marcos (a Hispanic serving institution). They found that over the seven years of data in their study, the cohorts of students in LA-supported courses had semester-over-semester decreases in DFW rates. The DFW rate fell from its historical rate of 25%–40% to below 20% in the final year of the study. They attributed the gradual decrease in DFW rate to faculty gradually becoming proficient at implementing LA-supported pedagogies and for the programmatic reforms to become institutionalized. These findings build on those of Alzen *et al.* [34] by expanding the set of contexts that LAs have been associated with improvements in DFW rates.

Researchers have identified other physics course features associated with improvements in student DFW rates. For example, research has associated decreases in physics course DFW rates with course transformations that feature small group and sense-making activities [55] and the use of 4-point grading scales instead of percentage grading scales [56]. Other researchers have focused on how student features, such as self-efficacy, can predict DFW rates.

For example, Sawtelle *et al.* [57] found that students with higher levels of select self-efficacy subconstructs passed introductory physics courses at higher rates than their peers.

Beyond the contexts of LAs and physics courses, some research investigated the relationship between supplemental instructors (SI) [58] and DFW rates. SI is a near-peer model where students who previously passed a course hold optional sessions in which they lead small groups working on testlike problems. Stanich *et al.* [59] examined a chemistry program that had expanded on the traditional SI support model to create a two-credit SI course that accompanied a traditional introductory chemistry course. They found mixed results on the association between the SI support course and DFW rates for underrepresented racial and ethnic minorities. After controlling for background and preparation variables, the students from underrepresented racial and ethnic groups had roughly the same DFW rates in both the control and SI groups. When controlling for who volunteered to be in the SI group, however, they found that the students from underrepresented racial and ethnic groups who were randomly accepted into the SI group may have had lower DFW rates than those in the non-SI group. The study lacked the statistical power to make any strong claims about differences in DFW rates across demographic groups.

In a large-scale investigation of DFW rates across the STEM disciplines at Northern Arizona University, Benford and Gess-Newsome [60] found that women had lower average DFW rates than men in every course they examined. They also examined race as a predictor for DFW rates. To have sufficient statistical power for their analysis, they aggregated all of the STEM courses. In their descriptive statistics, they found meaningful differences in DFW rates between racial groups. White students had the lowest DFW rates (22%) and Native American (43%) and African American (40%) students had the highest DFW rates. While there were differences in the descriptive statistics across demographic groups, they excluded both gender and race from their predictive models. The lack of student demographic variables in their models limited the ability of their findings to identify inequities. They found that student-centered courses, as measured by the research of teaching observation protocol [61], had lower overall DFW rates. While the findings of Benford and Gess-Newsome [60] were not specific to physics, they indicate that there are differences in DFW rates across course types and demographic groups.

These studies show that student-centered interventions can improve *overall* DFW rates for physics students. These studies also found consistent differences in DFW rates favoring white, female, continuing-generation students in college science courses and physics courses. They provide little guidance, however, on what interventions would improve the *equity* of DFW rates for students in physics courses. In the present study, we build upon this prior work by using a critical perspective to investigate DFW rates across demographic groups and across introductory physics courses that did and did not use LAs.

### III. CONCEPTUAL FRAMEWORK

#### A. Critical theory

Critical race theory (CRT) began in the 1970s and 1980s as a movement among U.S. legal scholars of varied racial backgrounds to address social injustices and racial oppression [62–64]. CRT explicitly assumes racism is ingrained in our institutional structures, focuses on the narratives and counternarratives of oppressed people, and identifies the importance of interest convergence between oppressed peoples and their oppressors in creating change [65,66]. The civil rights movement in the United States provides an example of interest convergence as it succeeded in part because segregationist policies undermined international diplomacy efforts during the cold war [67]. In the intervening years, CRT has been taken up by scholars in many fields, including education [68,69]. Each of these offshoots apply the defining characteristics of CRT, such as challenging the ideas of objectivity and claims of neutrality [62], in novel contexts. For example, disability CRT (DisCrit) [70–72] has been taken up by scholars to examine the intersection of racism and disabilities. To analyze and interpret our findings, we used a quantitative CRT (QuantCrit) [25,37,73] perspective.

#### B. QuantCrit

Critical research has historically used qualitative approaches to investigate the lived experiences of marginalized people and the social processes that create racist, sexist, and classist power structures [25,28,73]. QuantCrit emerged as a quantitative perspective [37] aligned with the core principles of critical research. QuantCrit complements qualitative studies by using large-scale data to represent student outcomes in ways that reveal structural inequities that reproduce injustices [37]. A QuantCrit perspective also pushes researchers to identify where society fails to measure the outcomes for marginalized groups, such as our societies failure to look at pregnancy outcomes generally and particularly for women of color [74]. Below, we describe three principles of QuantCrit [73] and the ways we strove to embody them in this investigation:

(1) *The centrality of oppression*.—We assumed that racism, sexism, and classism are complex and dynamic processes present throughout society that we must explicitly examine lest our statistical models legitimize existing inequities. Educational inequities come from hegemonic power structures creating educational and societal systems that cater to students from dominant groups. The continual marginalization of specific student populations creates educational debts [75]. These educational debts are not features of students, but debts that society owes marginalized students. Researchers can measure some aspects of educational debts with quantitative measures (e.g., representation, grades, test scores). However, quantitative measures cannot access all aspects of the educational debts owed by society nor can a single quantitative measure indicate that an intervention or institution has redressed all educational debts. These assumptions were at the heart of our work and informed each stage of our research.

(2) *Categories are neither natural nor given.*—All data are socially constructed and reflect the hegemonic power structures that created them. Grades, for example, are social constructs created by instructors and codified by our educational systems. How instructors assign grades is an idiosyncratic process that reflects the values and beliefs of individual instructors and the power structures of their discipline and university, rather than an abstract truth about a student.

Our models aggregate students by race, gender, and class. These categories do not represent any natural or scientific truth about students but are social constructs that maintain hegemonic power structures. The dynamic socially-negotiated natures of race, gender, and class does not diminish the very real effects of racism, sexism, and classism associated with them. We strive to clarify that our models are not measuring innate difference in students based on their race, gender, or first-generation (FG) status, but the impacts of multidimensional oppressive power structures on students marginalized by these social constructs. One way that we reflect this in our writing is through the explicit naming of racism, sexism, and classism in interpreting our models.

(3) *Data are not neutral and cannot speak for themself.*—We reject the idea that data is neutral and can speak for itself. Racist, sexist, and classist assumptions can shape every stage of collecting, analyzing, and interpreting data [26]. All of the demographic data used in our analysis reflected an institution's attempt to categorize and quantify aspects of student identities that fell along spectrums. For example, CSU Chico only allowed students to identify as a binary gender in their university paperwork. This practice marginalized students who identify as trans or gender nonconforming and limited our ability to examine some dimensions of oppression. The institutional classification of student racial identities and FG status were also socially constructed categories with their own sets of assumptions. More details on the demographic data are provided in Sec. V A. In analyzing this data, we used methods that we felt produced the most meaningful representation of the impacts of racism, sexism, and classism knowing that the data and methods were imperfect.

Some methods we used allowed us to create more inclusive and nuanced findings. For example, our use of Akaike information criterion corrected (AICc) to select our models and not using $p$ values to interpret them allowed us to model and discuss inequities in student outcomes that would have been lost using more traditional methods. Other methods, however, were necessary to develop our models but had clear limitations. For example, a small portion ($< 1\%$) of our students data had no response for gender. We did not know if this happened through an error in the campus systems or these students chose not to identify a gender on their university forms. It is possible that some of this missing data arose from trans and gender nonconforming students not having a nonbinary option and choosing not to answer the question. To account for this missing data, we included an additional gender variable (Gen.unknown) in our statistical models. Because the number of students without gender data was small, the model's predictions for the them are not meaningful and we do not discuss them in our analysis of our findings. This solution had advantages and drawbacks. The primary advantage was that it allowed us to include their data in the models without making any assumptions about their gender identities. A drawback of this solution is that the group's outcomes are excluded from our discussion of the findings. In creating and interpreting our models, we did our best to speak for the data in ways that identify injustices while acknowledging that our findings were shaped by our own imperfect methods.

To give voice to our findings, we operationalized equity in two competing ways (see Sec. III C) and used them to interpret our findings from multiple perspectives.

The underrepresentation of students from marginalized groups makes it difficult to collect large enough samples to investigate inequities with dependable statistical analyses. These challenges are exacerbated for studies that disaggregate across intersecting marginalized identities, such as for women of color. Some investigations incorrectly claim they found no differences across demographic groups because the analyses were underpowered and did not find differences with a $p$ value below 0.05. Lack of a statistically significant $p$ value should not be confused with lack of a meaningful effect [76–78]. QuantCrit researchers, in part, overcome this challenge by collecting large-scale datasets with enough statistical power to model the relationships between students' intersectional identities and their learning outcomes. While our work's foundations lie in the QuantCrit literature, we were informed by the prior work using intersectionality in physics [12–17], intersectionality in QuantCrit [25,26,28,79], and the foundational work in intersectionality [20]. This body of work was particularly informative for our statistical model building process. The recent emergence of large-scale databases of university science student data [51,80,81] has made it easier to get the statistical power needed to model the impacts of intersecting racist, sexist, and classist power structures. Even with these large-scale databases small samples for intersectional and underrepresented populations can obscure inequities. Rather than include $p$ values in our findings [82], we focus on transparency by reporting the point estimates and uncertainties from our models. This method prevents our results from focusing solely on groups well represented in the data but rather on differences that warrant attention.

### C. Operationalizing equity

Because data cannot speak for themself, we follow the advice of Rodriguez et al. [83] and Stage [37] and operationalized equity to interpret our findings from a social-justice perspective. Specifically, we offer two competing operationalizations of equity that we used to interpret our findings: (i) equality of outcomes and (ii) equity of individuality. We grounded these operationalizations of equity in the literature but we renamed them to ease the readers' interpretation and to align with Lee's [84] definition of equity and equality.

Equality of *outcomes* occurs when students from different demographic groups have the same average achievement at the end of a course regardless of their backgrounds. This perspective on equity has been called equity of parity [83,85] and equality on average across social groups [86]. This perspective takes a strong social-justice stance by placing the onus on the education system to allocate resources to eliminate inequalities and redress educational debts. In these scenarios, students from marginalized groups receive support that begins to repay their educational debts by overcoming the impacts of prior injustices.

Equity of *individuality* occurs when an intervention improves the outcomes of students from marginalized groups [83]. This perspective gets away from making comparisons with white, middle-class students and what Gutiérrez and Dixon-Román [87] refer to as "gap gazing." Rather, it focuses on research and interventions designed to advance the needs of individuals who are marginalized as a result of group identity. Gutiérrez [88] argues that the focus on achievement gaps supports a deficit model of students from marginalized groups. By only focusing on marginalized students, however, equity of individuality ignores the impact of interventions on students from dominant groups. By excluding students from dominant groups, equity of individuality may miss opportunities for interest convergence that promote equitable practices, thereby exacerbating the existing educational debts. For this reason, our examination of equity of individuality included the associations between LAs and outcomes for students from dominant groups.

### D. Positionality

Feminist theory has shown that all knowledge is marked by those who create it [89]. To be transparent about the position of the researchers in this work in relation to the power structures under investigation, we offer positionality statements [90] for each of the authors.

The following is the first author's, Ben Van Dusen, positionality statement. I identify as a White, cisgender, heterosexual, continuing-generation (CG) man with a color vision deficiency. I was raised in a pair of lower-income households but I now earn an upper-middle class income. I am an assistant professor and director of the LA program at the Hispanic serving institution where the study was performed. My experiences working with marginalized students, particularly those whom I have had the honor to mentor as LAs and as researchers, have motivated my attempts to use my position and privilege to dismantle oppressive power structures. As someone who seeks to be an ally it is easy to overlook my own privileges. I try to broaden my perspective through feedback from those with more diverse lived experiences than my own.

The following is the second author's, Jayson Nissen, positionality statement. My identity as a White, cisgendered, heterosexual, nondisabled man has provided me with power and opportunities denied to others in American society. I use my experience growing up in a poor home and as a veteran of the all-male submarine service to motivate reflecting on and working to dismantle my privilege. My work on this project was shaped by the post-positivist scientific traditions I was educated in and my activist goal to pursue scientific knowledge that can help identify and dismantle policies and systems of oppression. Because I am not a woman or a person of color and I now live in a higher income household, I brought a limited perspective to this work on racism, sexism, and classism.

To address potential homogeneity of author positionality and perspective, we elicited feedback from a diverse set of peers and employed Radical Copy [91] to perform an equity audit of the publication.

## IV. RESEARCH QUESTION

This study investigated the intersectionality of racism, sexism, and classism in physics courses. Specifically, we examined the associations between LA support in physics courses and DFW rates of marginalized student populations. To better understand these associations we addressed two research questions:

(1) To what extent are Learning Assistants (LAs) associated with decreases in DFW rates overall for students in introductory physics courses?

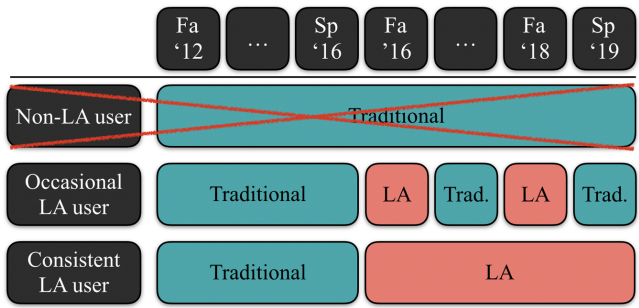

FIG. 1. The three types of instructors in the physics department. To control for instructors self-selection into using LAs, we only analyzed data for courses taught by the occasional and consistent LA-using faculty.

(2) To what extent are shifts in DFW rates associated with LA support in introductory physics courses mitigating the impacts of racism, sexism, and classism?

By investigating the association between LA support and DFW rates, the first research question addresses the extent to which LA programs can support more students succeeding in introductory physics courses. The second question addresses the extent to which LA support is associated with decreasing or increasing inequities. Even if the LA model reduces DFW rates, we should not consider it a success if it perpetuates existing racist, sexist, and classist oppressive structures. These findings will inform administrators and faculty seeking interventions that increase student success, persistence, and retention.

## V. METHODS

### A. Data collection

It was possible that the instructors who integrated LAs into their courses had different DFW rates than instructors who never integrated LAs into their courses. To control for potential self-selection bias between these groups of instructors, the analysis only included instructors who had taught the course with LA support for at least one semester (Fig. 1). We did not collect data from sections taught by instructors that never integrated LAs into any of their sections.

The data for the study came from 9 instructors who taught 41 sections of first semester introductory physics courses to 2312 students. The average DFW rate across all the courses was 26.1% (Table I). Forty-two percent of the data came from students in LA-supported courses. Five of the nine LA-using faculty in this study completed an online survey with open-ended questions about their use of LAs. The relevant questions on the survey were: Does having LAs change the way you teach your course(s)? If so, how? The responses were unanimous in stating that in the semesters they had LAs, the LAs allowed them to reduce the time spent lecturing and increase use of collaborative learning activities. The details of what and how collaborative learning activities were implemented varied by instructor.

The LA-supported physics courses typically had a ratio of 24 students/LA. The demographics of the LAs in our sample was not known, but the overall LA population at Chico State was similar to that of the student population. All of the LAs took the same LA pedagogy course during their first semester as LAs. The LA pedagogy course, however, was modified over the course of the investigation to include more activities focused on helping the LAs support marginalized student populations.

When collecting data on student race, CSU Chico only allowed students to select a single response from a list of groups that combined race and ethnicity (e.g., White, Black, Hispanic, and Asian). Our sample size limited our ability to disaggregate many of the race categories in our models. We combined all responses other than white into a single category labeled Black, Indigenous, and people of color (BIPOC). Within the BIPOC category, the responses included Hispanic (52%), Non-Resident Alien (12%), Unknown (12%), and Asian (11%), two or more races (8%), Black/African American (3%), American Indian/Alaskan Native (<1%), and Pacific Islander (<1%). We had no way to determine the race of students who selected Non-Resident Alien, unknown, or two or more races. The DFW rates for Non-Resident Alien students were moderately above the overall averages and for the unknown students were slightly below the overall averages.

The gender variable included three response categories: male, female, and unknown. The institution did not allow students to choose trans, gender nonconforming, nonbinary, and two-spirit students. The response categories, male and female, more closely reflect sex than gender. To address this mislabeling of student gender, we will use the terms men and women rather than male and female. The unknown responses were from students for whom the university had no gender identifier and made up less than 1% of the total responses. Some students who had missing gender data may have identified as trans or gender non-conforming chose not to select one of the binary gender options.

TABLE I. Descriptive statistics for the first semester courses in each of the two introductory physics sequences.

| Courses | | | | | Students | | | | | | | |
|---|---|---|---|---|---|---|---|---|---|---|---|---|
| | | | | | | Gender | | | Race | | FG status | |
| Math-basis | Instructors | Sections | DFW (%) | Total | LA supported (%) | Men (%) | Women (%) | Unknown (%) | White (%) | BIPOC (%) | FG (%) | CG (%) |
| Algebra | 3 | 9 | 20.9 | 862 | 74.8 | 57.4 | 41.9 | 0.7 | 46.5 | 53.5 | 50.1 | 49.9 |
| Calculus | 6 | 32 | 29.2 | 1450 | 22.4 | 84.4 | 14.6 | 1.0 | 39.7 | 60.3 | 43.1 | 56.9 |
| Total | 9 | 41 | 26.1 | 2312 | 42.0 | 74.4 | 24.8 | 0.9 | 42.2 | 57.8 | 45.7 | 54.3 |

TABLE II. Descriptive statistics disaggregated by demographics and instruction.

| Gender | Race | FG status | Instruction type | N | DFW rate(%) |
|---|---|---|---|---|---|
| Men | BIPOC | FG | Traditional | 316 | 37.7 |
| | | | LA | 218 | 28.9 |
| | | CG | Traditional | 316 | 42.1 |
| | | | LA | 123 | 17.9 |
| | White | FG | Traditional | 134 | 30.6 |
| | | | LA | 86 | 22.1 |
| | | CG | Traditional | 314 | 21.7 |
| | | | LA | 212 | 13.2 |
| Women | BIPOC | FG | Traditional | 86 | 26.7 |
| | | | LA | 135 | 17.8 |
| | | CG | Traditional | 73 | 27.4 |
| | | | LA | 69 | 10.1 |
| | White | FG | Traditional | 33 | 21.2 |
| | | | LA | 49 | 16.3 |
| | | CG | Traditional | 70 | 20.0 |
| | | | LA | 78 | 10.3 |
| Unknown | All | All | Traditional | 13 | 17.9 |
| | | | LA | 7 | 0 |

The university classifies students as first generation (FG) (as opposed to continuing generation [CG]) if they meet one of three criteria: (i) a student neither of whose biological or adoptive parents received a baccalaureate degree; (ii) a student who, prior to the age of 18, regularly resided with and received support from only one parent and whose supporting parent did not receive a baccalaureate degree; or (iii) an individual who, prior to the age of 18, did not regularly reside with or receive support from a natural or adoptive parent.

The data used in this study included 2312 students (Table I). Women made up 24.8% of the students, BIPOC made up 57.8% of the students, and FG students made up 45.7% of the students. The percentage of students who were BIPOC and/or FG was similar in the algebra- and calculus-based courses (Table I), but the algebra-based course had a much higher proportion of women (41.9%) than the calculus-based courses (14.6%). The descriptive statistics for each demographic group are included in Table II.

### B. Data analysis

To analyze the data, we generated hierarchical generalized linear models that predicted DFW outcomes while accounting for the nested structure of our dataset (students in sections) using the hglm [92] and lme4 [93] packages in RStudio v.3.5.1 [94]. Our hglm model parameters were fit using the extended quasilikelihood method. To determine which model was the best fit for our data, we used the dredge function in the MuMin package [95] to calculate the corrected Akaike information criterion (AICc) [96] for each combination of the variables. We used the model with the lowest AICc score as our final model. We used the AICc score, rather than variance explained, to select our final model for several reasons. AICc scores take into account the explanatory power of each variable without overly weighting model parsimony. While it is a common practice to use variance explained to select a final model (e.g., our own prior work [45,97]), using it to select models when investigating marginalized populations risks the models falling prey to a tyranny of the masses. Using predicted variance to select variables risks excluding marginalized students from groups with small representations in the data because the variance explained by a variable is proportional to its the sample size.

To examine whether the nested structure of our data necessitated the use of hierarchical models, we generated an unconditional model without predictor variables and calculated the intraclass correlation [98]. We found 8.6% of the variance at the course level; therefore, the best practice was to account for the hierarchical structure of the data in our model [97].

To generate our model for equity in DFW rates, we began by creating a model that included use of LAs as the only predictor variable so we could model the overall DFW rate in the introductory physics courses with and without LAs. We then explored using a level-2 course variable (use of LAs) and level-1 student demographic variables (race, gender, FG status) and the interaction effects between the student demographic variables with each other as well as with the use of LAs. Our examination of AICc identified the model that included some two-way interactions, but not three-way or four-way interactions as our best model of equity in DFW rates. The following equations describe the models of overall DFW rates and equity in DFW rates.

#### 1. Overall model

Level-1 equations (student level)

$$\text{DFW}_{ij} = \beta_{0j} + r_{ij}$$

Level-2 equations (course level)

$$\beta_{0j} = \gamma_{00} + \gamma_{01} \times \text{LA} + \mu_{0j}$$

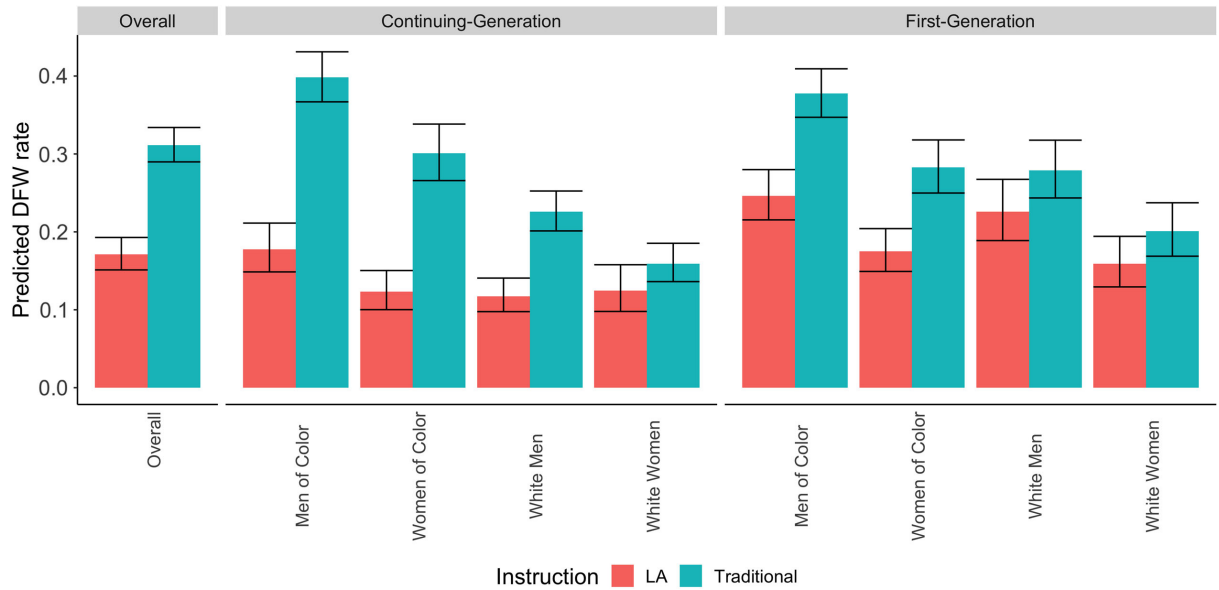

FIG. 2. Predicted DFW rates for each group of students with and without LAs, after accounting for instructor types. Error bars represent $+/-1$ S.E..

*2. Equity model*

Level-1 equations (student level)

$$\text{DFW}_{ij} = \beta_{0j} + \beta_{1j} \times \text{Woman} + \beta_{2j} \times \text{Gen.unknown} + \beta_{3j} \times \text{BIPOC} + \beta_{4j} \times \text{FG} + \beta_{5j} \times \text{BIPOC} \times \text{FG} + r_{ij}$$

Level-2 equations (course-level)

$$\beta_{0j} = \gamma_{00} + \gamma_{01} \times \text{LA} + \mu_{0j}$$
$$\beta_{1j} = \gamma_{10}$$
$$\beta_{2j} = \gamma_{20}$$
$$\beta_{3j} = \gamma_{30} + \gamma_{31} \times \text{LA}$$
$$\beta_{4j} = \gamma_{40} + \gamma_{41} \times \text{LA}$$
$$\beta_{5j} = \gamma_{50}$$

To help interpret the uncertainty around the predictions of our models, we included standard error values for each coefficient. We do not include $p$ values because of their consistent misuse in the sciences [76,78] and in research on equity [99]. Using $p$ values to examine equity is, in part, problematic because scientists often interpret them as go no-go tests. Since $p$ values are sample size dependent, they can show that large and meaningful effects for small groups of students are not statistically significant. Scientists and science consumers often think a result that is not statistically significant is not meaningful, but this is incorrect [78].

The statistical packages we used to generate the predicted DFW rates for each demographic group (Table III and Fig. 2) in the investigation of equity of outcomes allowed us to easily calculate the uncertainty for each estimate using the standard errors. However, our investigation of equity of individuality examined the difference between predicted DFW rates (Fig. 3). This simple subtraction complicated the calculation of standard errors. To address this challenge, we used bootstrapping [100] with 1000 subsamples resampled at the section level to estimate these means and standard errors for the differences between student DFW rates in LA and traditional courses.

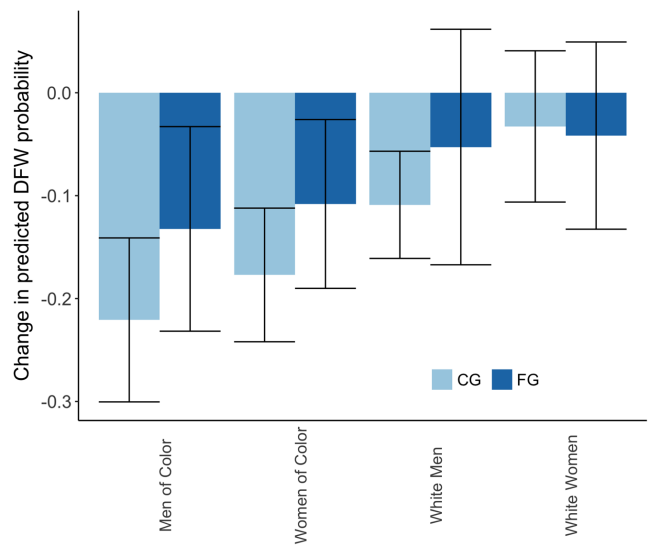

FIG. 3. Predicted DFW rate decreases for each demographic group in LA-supported courses versus traditional courses, accounting for instructor types. Error bars represent 95% confidence intervals.

TABLE III. Predicted DFW probabilities by student demographic, accounting for instructors.

| Gender | Race | FG Status | Inst. | DFW Rate (%) | One S.E. Range (%) |
|---|---|---|---|---|---|
| Men | BIPOC | FG | Trad. | 37.8 | 34.7–40.9 |
| | | | LA | 24.6 | 21.5–28.0 |
| | | CG | Trad. | 39.9 | 36.7–43.1 |
| | | | LA | 17.8 | 14.9–21.1 |
| | White | FG | Trad. | 27.9 | 24.4–31.8 |
| | | | LA | 22.6 | 18.9–26.7 |
| | | CG | Trad. | 22.6 | 20.1–25.2 |
| | | | LA | 11.7 | 9.8–14.1 |
| Women | BIPOC | FG | Trad. | 28.2 | 25.0–31.8 |
| | | | LA | 17.5 | 14.9–20.4 |
| | | CG | Trad. | 30.1 | 26.6–33.8 |
| | | | LA | 12.3 | 10.0–15.0 |
| | White | FG | Trad. | 20.1 | 18.9–23.7 |
| | | | LA | 15.9 | 12.9–19.4 |
| | | CG | Trad. | 15.9 | 13.6–18.5 |
| | | | LA | 12.5 | 9.8–15.8 |

## VI. FINDINGS

The findings section presents the outputs of DFW models and our interpretation of those findings for equality of outcomes and equity of individuality. The section begins with a table of the coefficients for the two DFW models (Table IV). The coefficients are logits that are not easily interpretable, particularly in isolation. To help make sense of the model outputs, we included plots of the model's predicted DFW rates for each group (Fig. 2) and the difference between a group in LA versus traditional courses (Fig. 3).

TABLE IV. Model for equity in DFW rates, accounting for instructor types.

| | Overall model | | Equity model | |
|---|---|---|---|---|
| Variable | Coefficient | S.E. | Coefficient | S.E. |
| Intercept, $\beta_{0j}$ | | | | |
| Intercept, $\gamma_{00}$ | −0.793 | 0.103 | −1.232 | 0.146 |
| LA, $\gamma_{01}$ | −0.785 | 0.179 | −0.785 | 0.241 |
| Woman, $\beta_{1j}$ | | | | |
| Intercept, $\gamma_{10}$ | ⋯ | ⋯ | −0.432 | 0.128 |
| Gen.unknown, $\beta_{2j}$ | | | | |
| Intercept, $\gamma_{20}$ | ⋯ | ⋯ | −1.059 | 0.646 |
| BIPOC, $\beta_{3j}$ | | | | |
| Intercept, $\gamma_{30}$ | ⋯ | ⋯ | 0.821 | 0.155 |
| LA, $\gamma_{31}$ | ⋯ | ⋯ | −0.336 | 0.227 |
| FG, $\beta_{4j}$ | | | | |
| Intercept, $\gamma_{40}$ | ⋯ | ⋯ | 0.283 | 0.191 |
| LA, $\gamma_{41}$ | ⋯ | ⋯ | 0.501 | 0.221 |
| BIPOC*FG, $\beta_{5j}$ | | | | |
| Intercept, $\gamma_{50}$ | ⋯ | ⋯ | −0.371 | 0.216 |

### A. LAs and overall DFW rates

Our first research question was: to what extent are learning assistants (LAs) associated with decreases in DFW rates overall for students in introductory physics courses? The overall model shows a strong association between LAs and lowered DFW rates. When taught by the same instructor, courses in the introductory physics sequence have predicted DFW rates of 31.1% without LAs and 17.1% with LAs (Fig. 2). This difference was much larger than the uncertainty in the measurements.

### B. LAs and equality of outcomes

Our second research question was: to what extent are shifts in DFW rates associated with LA support in introductory physics courses mitigating the impacts of racism, sexism, and classism? To answer that question we defined equity in two ways. The first definition was equality of outcomes, in which we compared the differences in predicted DFW rates for each demographic group in traditional courses and again in LA-supported courses. The model for equity in DFW rates showed meaningful differences in predicted DFW rates across demographic groups in both contexts (Fig. 2). In traditional courses, predicted DFW rates ranged from 40% (CG men of color) to 16% (CG White women). In LA-using courses, predicted DFW rates ranged from 25% (FG men of color) to 12% (CG White women). Regardless of course type, being a man, a BIPOC, or a FG student were all associated with higher predicted DFW rates. With only the exception of FG White men in LA-supported courses, BIPOC students had the highest predicted DFW rates regardless of FG status or course type. As we discuss below, these results contrasted some of our expectations based on the role of hegemonic power structures in educational debts. First, women were less likely than men to receive DFW grades no matter the type of instruction. Second, the relationship between FG status and DFW grades varied between White students and BIPOC students across instructional types. The data did not indicate an additional education debt for FG men and women of color in traditional courses. The differences in predicted DFW rates were greatly decreased in LA-supported courses. For example, CG women of color, White men, and White women all had predicted DFW rates that ranged from 12%–13% in LA supported courses. While there was meaningful progress toward equity of outcomes, some inequities persisted in LA-supported courses showing that equity of outcomes did not occur in either course context.

### C. LAs and equity of individuality

To answer our second research question from another perspective we used a second definition of equity, which we called equity of individuality. To determine if equity of individuality occurred in LA-supported courses, we compared the differences in predicted DFW rates for each demographic group in LA-supported courses versus themselves in traditional courses. Figure 3 shows the predicted DFW rates decreased by meaningful amounts in LA-supported courses for each demographic group. However, the smaller decrease for FG White students and larger uncertainty in the measurement due to smaller samples means that we cannot be completely confident in this difference. The 95% confidence intervals for FG White men, FG White women, and CG White women all include positive values, unlike the other predicted values. The largest decreases in predicted DFW rates were for BIPOC students. CG students had larger predicted decreases in DFW rates than FG students, with the exception of White women. The decreases in predicted DFW rates for each marginalized student group means the LA-supported courses achieved equity of individuality.

## VII. DISCUSSION

The models identified educational debts incurred by racist and classist power structures impacting student outcomes in both traditional and LA-supported courses. However, these models did not identify inequities in DFW rates against women. The equity of individuality analysis, however, predicted lower DFW rates, up to 22 percentage points, for students from marginalized groups in LA-supported courses. This large decrease in DFW rates across all groups indicated the integration of LAs into these physics courses achieved an equity of individuality. In terms of passing the course, students from marginalized groups have better outcomes in LA-supported courses.

The decreases in DFW rates associated with LAs were not equal across all marginalized groups. The largest predicted decreases in DFW rates in LA-supported courses occurred for BIPOC independent of gender (Fig. 3). First-generation (FG) students, however, had smaller predicted decreases in DFW rates associated with LAs than their CG peers. These results indicate adding LA support to a course may not be sufficient to address the role of classism in student outcomes.

The equity model also found an interaction between FG and BIPOC variables indicating a larger decrease in DFW rates for FG BIPOC students than a model not accounting for these interactions would have predicted. This highlights the importance of taking an intersectional approach that does not assume that the impacts of racism, sexism, and classism are additive [79]. White and BIPOC students likely experience classism differently in their educational experience. FG status may also act differently as a proxy of class for White and BIPOC students. Understanding the role of classism, and how to investigate it, in physics education will take a concerted effort from our community. Class is a taboo subject in the United States. We must protect students privacy and their right to informed consent. However, we must also make sure that these values are not co-opted by education systems to hide injustices.

While the equity model associated LA support with equity of individuality (i.e., improved outcomes for marginalized students), they did not associate LA support with equality of outcomes (i.e., equal outcomes across demographic groups). For example, the equity model predicted 25% of FG men of color received DFW grades in LA-supported courses while only 12% of CG White women received DFW grades. While the model predicted lower DFW rates in LA-supported courses, the inequities identified educational debts due to racism and classism in student failure rates in LA-supported courses. The lower DFW rate for women indicates these courses did not increase educational debts due to sexism.

A major difference between sexism and racism and classism is that Title IX of the Education Amendments of 1972 legally protects women against discrimination. No similar legal protections exist across races and classes. Grades provide an accessible metric of systemic sexism within a course, program, or institution. Institutional and social pressures around sexism may mean that sexism had a smaller role in these physics courses or it may mean that grades no longer reflect the sexism women experienced in these courses. For example, Seymour [9] found that women were highly capable but the culture of many science and engineering courses (e.g., competitive, cutthroat, and hostile) pushed women to leave STEM degrees. Research also shows many women experience microaggressions and blatant sexism in their physics education [10,11], and that stereotypical environments' lack of representation of women undermine women's sense of belonging and performance in STEM learning environments [101]. In their review of the literature on gender differences in representation across STEM discipline, Cheryan et al. [102] found that masculine cultures, lack of early experiences for girls, and gender differences in self-efficacy explain the large gender differences in participation in physics, engineering, and computer science. The difference in gender representation between the algebra- and calculus-based courses illustrate these differences in participation across STEM fields within the data used in this study. Better understanding the way students experience sexism, racism, and classism in introductory physics courses and the role of LA programs in mitigating those experiences requires further study with a diversity of methods. A single metric and a single study is insufficient for claiming the elimination of sexism in a course when that course is situated in a field and a society with a long history of sexism.

Because of a lack of data on prior preparation, our models did not account for educational debts that students brought into the physics courses. We have no reason to believe, however, there were differences in the educational debts between students in the LA-supported and traditional sections at the start of the semester.

Without data on prior preparation we cannot account for the role of sexist power structures beyond the course and institutions in affecting the educational debt women left the course with.

## VIII. CONCLUSIONS AND IMPLICATIONS

The decreased DFW rate in LA-supported course raises questions about how LA support relates to the lower DFW rates. The courses with LAs could have given more passing grades because they used relaxed grading criteria or used tests that gave higher average scores. We attempted to mitigate this possibility by excluding instructors who had never used LAs in the course. In their responses to a question about their grading practices in the online survey administered to the instructors, none reported making shifts in their grading criteria or assessments in the semester their courses were LA supported. Future work can test the possibility that DFW rates improved because of more lenient grading by comparing the grade structures and task difficulties when instructors do and do not have LAs. A second possibility is that the students learned more in the courses with LAs. This is supported by the instructors reporting using more student-centered collaborative teaching techniques when their courses included LAs. These teaching techniques support greater conceptual learning than teacher-centered lecture [45,103,104]. Van Dusen and Nissen [45] found that collaborative instructional practices increased conceptual learning across all demographic groups but that it did not eliminate differences across demographic groups. Greater conceptual learning in the LA-supported courses would have resulted in higher grades if the grading practices were similar between traditional and LA-supported courses. Future work can test the possibility that DFW rates improved because of increased learning by comparing student concept inventory data when instructors do and do not have LAs.

The decreases in DFW rates associated with LAs are meaningful and reduce inequities in physics student outcomes. The models associated LA support with lower DFW rates for all groups of students and the largest decreases occurred for BIPOC students. While LA support lead to equity of individuality, it did not eliminate differences across demographic groups. LA-supported courses did not achieve equality of outcomes. Achieving equality of outcomes through a single intervention is unlikely because it requires not only eliminating the impacts of racism, sexism, and classism within the course but also mitigating the impacts of educational debts incurred before and outside of the course. This LA program did not eliminate inequities but was associated with mitigation of some educational debts owed marginalized students. Continued investigations within this LA program can aid in further addressing these injustices.

Within each gender and racial group the FG students had smaller decreases in DFW rates associated with LAs. Our data cannot speak to whether these inequities arise from differences in the needs and resources of FG students, how LAs and FG students interact, or through other mechanisms. We are not aware of any large-scale investigations of the experiences of FG physics students that could inform our conclusions.

LA programs represent an appealing tool to improve equity on a large scale because LA programs create an interest convergence between marginalized students and those with power. Interest convergence in CRT [66] argues that improved outcomes for marginalized students is not sufficient motivation to create change within existing power structures. Only when those with power see a program as benefiting them or their interest groups will the program garner the resources to scale and sustain it. As both students from historically marginalized communities and students from dominant identity groups performed better in LA-supported courses and LA programs offer prestige for an institution, LA programs are well positioned to solicit support from those with the power to fund them. The effects of this interest convergence are illustrated by over 300 institutions joining the LA Alliance. The LA model has the potential for widespread improvement of equity in STEM courses.

There is a danger, however, that LA programs could increase systemic inequities. LA programs require resources including paying LAs and faculty and staff time and expertise to run the program and teach the pedagogy course. The institutions with the greatest resources tend to primarily serve White, middle-upper class students. If well resourced institutions disproportionately adopt LA programs, those programs will perpetuate existing racist and classist power structures by disproportionately benefiting White, middle-upper class students. For LA programs to counteract injustices at the national level they need support at institutions serving marginalized students, such as Hispanic serving institutions, minority serving institutions, historically black colleges and universities, and two-year colleges.

## IX. LIMITATIONS AND FUTURE RESEARCH

This study only represents outcomes for students in introductory physics at a single institution using a single type of course transformation. Increasing the scale of this work to include other near-peer programs (e.g., supplemental instruction), physics courses, STEM disciplines, and institutions will inform the extent of inequities across institutions and disciplines and the extent to which different course transformations address those inequities. Accounting for differences in prior preparation across institutions and course types can reveal oppressive power structures may reveal patterns of oppression hidden within data from a single institution. The study also lacked the statistical power to disaggregate across racial groups. Most BIPOC in this study self-identified as Hispanic. The impact of racism can vary widely across racial groups and settings [26,28,45]. Collecting data spanning course and institutional contexts with enough statistical power to disaggregate the data across demographic groups requires multi-institution collaborations. Large-scale studies can identify the impacts of oppressive power structures across institutions and can provide educators and administrators with guidance relevant to their context.

Our findings also highlight the need for further investigation into the causes of LA support having smaller associated decreases in DFW rates for FG students. To identify the underlying causes of these inequities will likely require the inclusion of qualitative methods. One potential path forward for this research would be to perform a mixed methods investigation of FG student outcomes that fuses quantitative data and testimonios, as demonstrated by Covarrubias et al. [26]. The quantitative data can provide the statistical power for a large-scale QuantCrit analysis and the testimonies can provide the rich descriptions of students' lived experiences.

## ACKNOWLEDGMENTS

This work was funded in part by NSF-IUSE Grants No. DUE-1525338 and No. 1928596 and is contribution No. LAA-065 of the Learning Assistant Alliance. We would like to thank McKensie Mack for leading the equity audit on this manuscript. We would also like to thank Tray Robinson and Ian Her Many Horses for their feedback on our work. We acknowledge and are mindful that the location where our research was performed stands on lands that were originally occupied by the first people of this area, the Mechoopda, and we recognize their distinctive spiritual relationship with this land and the waters that run through the campus.